# Moiré with flat bands is different

Tero T. Heikkilä (University of Jyvaskyla, Department of Physics and Nanoscience Center, P.O. Box 35 (YFL), FI-40014 University of Jyväskylä, Finland) and
Timo Hyart (International Research Centre MagTop, Institute of Physics, Polish Academy of Sciences, Aleja Lotnikow 32/46, PL-02668 Warsaw, Poland)

Recent experimental discoveries of superconductivity and other exotic electronic states in twisted bilayer graphene (TBG) call for a reconsideration of our traditional theories of these states, usually based on the assumption of the presence of a Fermi surface. Here we show how such developments may even help us finding mechanisms of increasing the critical temperature of superconductivity towards the room temperature.

Last year's **Physics World** *Breakthrough of the Year* award went to the experimental discovery[1] of an exotic superconducting state in a system where two layers of graphene were twisted close to a "magic" angle of around 1 degree. This angle was believed to be important because of an earlier prediction of a "flat" electronic band taking place at this angle.[2] In flat bands the density of electronic states is high, paving the way for strong interaction effects. Soon after the first discovery, another group reported similar type of behavior in their samples,[3] and showed that the superconductivity can be strongly affected by applying hydrostatic pressure. Both groups also managed to switch superconductivity on and off by an *in situ* tuning of the electronic density via a nearby gate voltage.

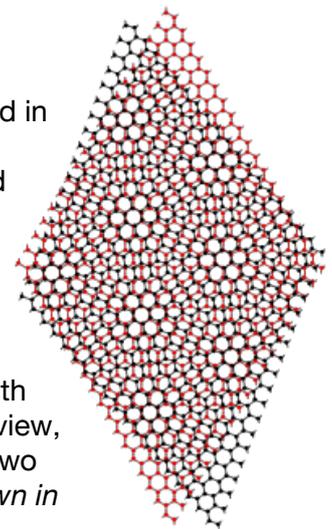

*Figure 1: Moiré pattern formed by two twisted honeycomb lattices.*

However, superconductivity was not the only type of exotic effect discovered in such systems: during the past year different groups have reported measurements of a correlated insulator state[4] tunable by a gate voltage, and a ferromagnetic state at a certain value of the electronic density.[5] Because of the closeness to the insulating state, the superconducting state was compared to that found in high-temperature superconductors, although TBG became superconducting only at a temperature around one Kelvin.

Needless to say, these observations lead to a flurry of theoretical activity, with dozens of papers trying to explain the observations from different points of view, including our own.[6] This quest is still on-going, and here we try to illustrate two fundamental aspects of these systems. First, *we cannot extend what is known in regular Fermi surface systems to those with flat bands by simply increasing the density of states or considering some generic low-energy Hamiltonian.* Rather, in many cases we need to know the structure of the entire flat band. Secondly, we discuss *how it might be possible to control the critical temperature of superconductivity in these systems.*

Let us start by sketching why the flat band physics becomes essential in TBG. Placing two honeycomb lattices on top of each other and twisting them relative to each other (Fig. 1) leads to the formation of a moiré pattern that looks periodic, such that the period increases as the twist angle becomes smaller. The strict periodicity holds only for certain commensurate angles,

---

otherwise the moiré lattice is quasiperiodic. However, as a first approximation we may assume that the electronic response is a smooth function of the angle, and concentrate on studying the commensurate angles. Periodicity enforced by the interlayer coupling means that we are allowed to use Bloch's theorem, and describe the electronic spectrum within the Brillouin zones of the superlattice. This is obviously not yet enough for the flat band formation, but we need two more concepts for it. First, graphene is a semimetal with a Dirac-point spectrum around two valleys in momentum space. Second, because of the twisting the valleys are shifted in momentum space with respect to each other (see Fig. 2(a)). For uncoupled layers, the energy bands cross between the two Dirac points (Fig. 2(b)). Coupling turns the crossing into an avoided crossing, hybridizing the modes from the two layers. Increasing the coupling moves one of the hybridized levels closer to zero energy (Fig. 2(c-d)), and at a critical value of the coupling the energy of this level ceases to depend on momentum, and the band becomes flat. This critical value of the coupling depends on the distance between the valleys, which is a function of the twist angle. This is how we can understand not only the formation of the (approximate) flat band, but also the fact that the magic angle can be tuned with pressure: applying pressure moves the layers towards each other, and therefore affects the interlayer coupling.

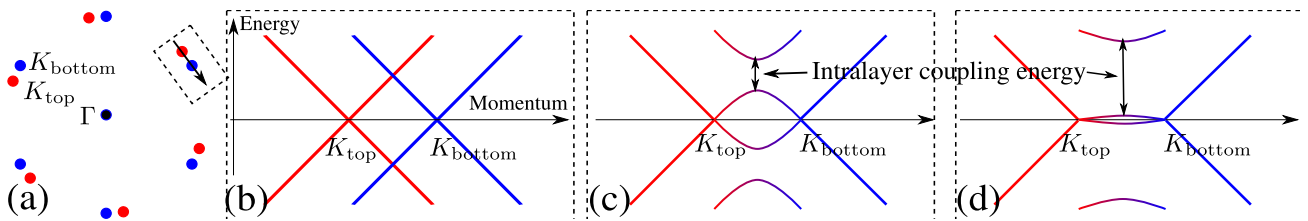

*Figure 2: Sketch of the flat band formation. (a,b): Dirac points of the two uncoupled but rotated layers form at slightly different points in the momentum space. (c): coupling the layers hybridizes the levels. (d): at a critical value of the coupling or of the distance between the two valleys ($K_{top}$ and $K_{bottom}$), the hybridized band becomes flat.*

The picture given here is a schematic one, but can be reproduced by more microscopic calculations. For example, Fig. 3 shows the spectrum calculated in Ref. 6 around the magic angle. It shows the formation of a pair of (spin degenerate) flat bands spanning the first Brillouin zone of the superlattice around each graphene valley. In the experiment the relevant energy scale for superconductivity is determined by the critical temperature, which means that one should zoom in to the range of the order of 1 meV (Fig. 3(c)), where even the band at the "magic angle" (here 0.96°, precise value depends on the employed model) ceases to be entirely flat.

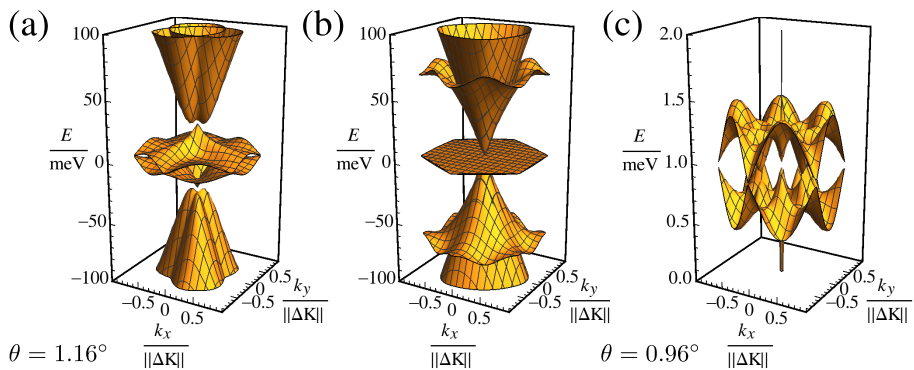

*Figure 3: TBG spectra from a tight-binding calculation along with a Dirac point approximation, see Ref. 6. (a) Above and (b,c) at the magic angle.*

Now, how is this different compared to the ordinary systems with near-quadratic dispersions, where majority of our knowledge of the superconducting state derives from? If the Fermi energy is far from the region with the flat bands, the difference is minor. But close to half-filling (Fermi energy close to zero), the picture is different and schematized in Fig. 4 for three different types of spectra: ordinary quadratic spectrum with a Fermi surface at some finite energy, Dirac (or Weyl) spectrum, and the flat band. The same figure also shows how the critical temperature for superconductivity scales with the superconducting coupling constant in the three cases. The "ordinary" case is the regular Bardeen-Cooper-Schrieffer (BCS) theory[7] whose microscopic coupling mechanism based on electron-phonon coupling was explored by Eliashberg.[8] It showed how the critical temperature is given by the Debye temperature times an exponentially small factor containing the electron-phonon coupling constant and the density of states at the Fermi level. Later it was also shown how the direct electron-electron interactions can be included via a pseudopotential that is often treated semi-phenomenologically.[9] This approach works well for electron-phonon mediated superconductivity in systems with Fermi surfaces.

One can also generalize the BCS theory for other types of electronic spectra. For Dirac or Weyl semimetals the density of states becomes very small close to the Dirac or Weyl points. On the mean-field level, this means that interaction effects are suppressed. For example, the Cooper instability according to which any Fermi surface is unstable to an infinitesimally small attractive interaction is turned into a more stringent condition for superconductivity: at the Dirac or Weyl points, superconducting state is obtained only above a critical coupling strength.[10]

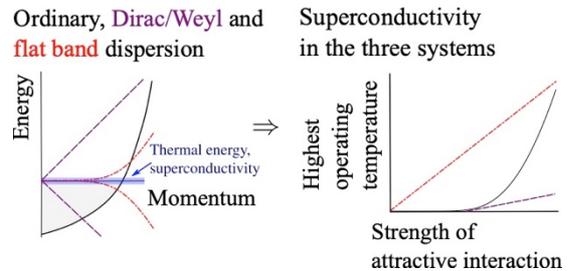

Figure 4: Sketches of energy spectra in three different generic types of electron systems, along with the behavior of the mean-field superconducting critical temperature in the three systems. The shaded region shows the typical energy ranges around the Fermi level.

If ordinary Dirac or Weyl fermions do not easily become superconducting, the opposite is true for flat bands: unlike in the usual Eliashberg theory, the (mean-field) interaction driven broken-symmetry phase has a critical temperature that is a linear function of the interaction strength.[11] It can thus be much larger than in conventional superconductors – perhaps up to room temperature? In TBG, the superconducting energy scales tend to be comparable to the bandwidth of the approximate flat band and thus approach this flat-band limit.[6] This means that almost all theoretical results we know based on Fermi surfaces cease to work in such systems. In particular, whereas typically many physical quantities especially related to electron transport can be obtained by concentrating on the Fermi surface or the vicinity of Dirac or Weyl points, in this case we have to take into account the contribution from the whole band, including for example effects originating from the quantum metric of the Bloch functions.[12] Another interesting feature of the flat-band systems is that they are not only unstable against the formation of the superconducting state in the case of an infinitesimally small attractive interaction but they are also susceptible for the formation of other types of correlated states (e.g. spin, charge or orbital order) in the case of infinitesimally small repulsive interaction.[11] This gives an intuitive explanation why so many different correlated states have been observed in moiré superlattices.

The experimentally obtained critical temperatures for the twisted bilayer graphene (TBG) are of the order of one or two kelvin at maximum. Based on the electron-phonon model (assuming it is the valid approach, which is still under debate), how can we understand this low value? It is not due to the weakness of the electron-phonon interaction, but rather from the small size of the flat band: In twisted bilayer graphene, the flat band takes place in the first Brillouin zone of the moiré superlattice. In momentum space the area of the flat band is inversely proportional to the square of the superlattice lattice constant, and the latter is quite large around the magic twist angles. To increase the critical temperature, we would hence need a way to create more extended flat bands.

---

[7] J. Bardeen, L. N. Cooper, and J. R. Schrieffer, "Microscopic Theory of Superconductivity", Phys. Rev. **106**, 162 - 164 (1957).
[8] G. M. Eliashberg, Sov. Phys. JETP **11**, 696 (1960).
[9] P. Morel, and P. Anderson, Phys. Rev. **125**, 1263 (1962).
[10] N.B. Kopnin and E.B. Sonin, Phys. Rev. Lett. **100**, 246808 (2008).
[11] N. B. Kopnin, T. T. Heikkilä, and G. E. Volovik, Phys. Rev. B **83**, 220503 (2011). See also R. Ojajärvi, T. Hyart, M.A. Silaev, and T.T. Heikkilä, Phys. Rev. B **98**, 054515 (2018) that generalizes the Eliashberg theory, along with the pseudopotential corrections, to the case of flat bands.
[12] For example, S. Peotta and P. Törmä, Nature Commun. **6**, 8944 (2015).

From the viewpoint of the theory more extended flat bands can be obtained in multilayer systems. The extreme case would be bulk graphite where it is known that extended flat bands appear in systems containing interfaces between differently twisted or stacked graphite regions.[13] Experimental signatures of high-temperature superconductivity have been reported in such systems,[14] but the community has not widely accepted these results because of the strong sample dependence of the results and the fact that they have not been reproduced by other experimental groups. The recent experimental advances have already allowed creating multilayer systems with tunable twist angles and it would be interesting to see a systematic study how the extension of the flat band in the momentum space affects the critical temperature. Well-controlled repeatable experiments in multilayer systems could also shed new light on the earlier experiments and lead to a better understanding of the limits of the critical temperature in these systems. In any case, understanding the characteristics of this type of exotic electron systems also requires a fresh look on the "well-known" theories.

---

[13] T.T. Heikkilä and G.E. Volovik, Chapter 6 in the book "Basic Physics of functionalized Graphite" (Edited by Pablo Esquinazi, Springer 2016) [arXiv:1504.05824]; T. Hyart and T. T. Heikkilä, Phys. Rev. B **93**, 235147 (2016).

[14] P. Esquinazi, Pap. Phys. **5**, 050007 (2013).